   \documentstyle[aps,prl,twocolumn,epsfig]{revtex}

\begin{document}
   \draft

   \twocolumn[\hsize\textwidth\columnwidth\hsize\csname @twocolumnfalse\endcsname%

   \title{L\'{e}vy statistics in coding and non-coding nucleotide sequences}
   \author{Nicola Scafetta$^{1}$, Vito Latora$^{2,3}$, 
   and Paolo Grigolini$^{1,4,5}$.}
   \address{$^{1}$Center for Nonlinear Science, University of North Texas,
   P.O. Box 311427, Denton, Texas 76203-1427 }
   \address{$^{2}$ Dipartimento di Fisica e Astronomia, Universit\'a di 
   Catania, and INFN, Corso Italia 57, 95129 Catania, Italy}
   \address{$^{3}$Laboratoire de Physique Th\'eorique et Mod\'eles 
   Statistiques,
   Universit\'e Paris-Sud, Bat. 100, 91405 Orsay Cedex, France}
   \address{$^{4}$Dipartimento di Fisica dell'Universit\`a di Pisa and
   INFM, Piazza Torricelli 2, 56127 Pisa, Italy}
   \address{$^{5}$Istituto di Biofisica CNR, Area della Ricerca di Pisa,
   Via Alfieri 1, San Cataldo 56010 Ghezzano-Pisa, Italy}
   \date{\today}
   \maketitle

   \begin{abstract}
   We propose a new method of statistical analysis of 
   nucleotide sequences yielding the true scaling without 
   requiring any form of de-trending. With the help of 
   artificial sequences that are proved to be statistically 
   equivalent to the real DNA sequences we find that 
   power-law correlations are present in both coding and 
   non-coding sequences, in accordance with the recent work 
   of other authors. We also afford a compelling evidence
   that these long-range correlations generate L\'{e}vy 
   statistics in both types of sequences.

   \end{abstract}
   \vspace{0.5cm}
   %
   ] 
   %


   The recent progress in experimental techniques of molecular genetics 
   has made available a wealth of genome data (see for example the 
   NCBI's Gen-Bank data base of Ref. \cite{genbank}). 
   This has triggered a large interest 
   in both the mechanics of folding \cite{torcini} and 
   the statistical analysis of DNA sequences. 
   This latter aspect, of interest for the present 
   letter, has been discussed by many authors 
   \cite{stanley1,wli,voss,stanley2}. These pioneer 
   papers mainly focused on the controversial issue of whether 
   long-range correlations are a property shared by both coding and 
   non-coding sequences or are only present in non-coding sequences. 
   The results of more recent papers \cite{mohanti,audit} 
   yield the convincing conclusion that the former condition applies. 
   However, some statistical aspects of the DNA sequences are 
   still obscure, and it is not yet known to what extent the 
   dynamic approach to DNA sequences proposed by the 
   authors of Ref. \cite{barbi} is a reliable picture 
   for both coding and non-coding sequences. 
   The later work of Refs. \cite{paolophdthesis} and 
   \cite{buiattino} established a close connection between long-range 
   correlations and the emergence of non-Gaussian statistics, confirmed 
   by Mohanti and Narayana Rao \cite{mohanti}. 
   However, according to the dynamic approach of 
   Refs. \cite{barbi,allegro} this non-Gaussian statistics should be L\'{e}vy, 
   and this aspect has not yet been assessed with compelling evidence.

   In this letter we propose a new technique of statistical analysis, 
   the Diffusion Entropy (DE) method, 
   and we prove that the joint use of this new technique and 
   of the Detrended Fluctuation Analysis (DFA), 
   applied to DNA sequences by the authors of Ref. \cite{stanley3}, 
   allows us to: 

   1) establish the presence of long-range correlations 
   in coding as well as in non-coding sequence;

   2) assess the L\'{e}vy nature of the resulting non-Gaussian 
   statistics. 

   In particular we analyze the two DNA sequences studied 
   in Ref. \cite{stanley3}. 
   These two sequences are the human T-cell receptor alpha/delta locus, 
   Gen-Bank name HUMTCRADCV, a non-coding 
   cromosomal fragment of $M = 97630$ bases (composed of less than 10\% of 
   coding regions), 
   and the Escherichia Coli K12, Gen-Bank name ECO110K, 
   a genomic fragment with $M = 111401$ bases 
   consisting of mostly coding regions (it contains more that 80\% of 
   coding regions). 
   We build up a random walk trajectory in the $x$-space with 
   the following prescription \cite{stanley2}. 
   The site position $t$ is interpreted as ``time''. 
   The walker $x(t)= x(t-1) + \xi(t)$ takes a step up [$\xi(t) = +1$]
   for each pyrimidine at position $t$, 
   and a step down [$\xi(t) = -1$] for each purine.
   Thus a DNA sequence becomes equivalent to a single trajectory 
   from which we have to derive many distinct trajectories as we shall 
   show below.
   The basic tenet of many techniques, currently used to analyze time 
   series, is the detection of scaling\cite{mandelbrot,feders}. Scaling 
   is a property of diffusion processes where reference to the 
   same distribution form can be done by 
   relating the space variable $x$ to the time variable $t$ via the key 
   relation:
   \begin{equation}
   x \propto t^{H}.
   \label{scalingdefinition}
   \end{equation}

   Ordinary Brownian motion has a time auto-correlation 
   function $\Phi_{\xi}(t)$ equal to zero, 
   except for $\Phi_{\xi}(0)=1$ , and is known to yield $H=1/2$. 
   The detection of $H \neq 1/2$ implies instead the presence of 
   extended correlation, i.e. a correlation function $\Phi_{\xi}(t)$ 
   described by a power law, which, in turn, can be interpreted as 
   a signature of the complex nature of the observed process.
   The detection of the true scaling, however, often involves the 
   adoption of detrending procedures, since a steady 
   bias hidden in the data produces effects which might be 
   mistaken for a striking departure from Brownian diffusion, 
   while the interesting form of scaling must be of 
   totally statistical nature.
   In the case of the DNA walk, the different trajectories of the 
   diffusion process are generated in the following way. 
   For each time $t$ we can construct $M-t+1$ trajectories 
   of length $t$: 
   \begin{equation}
   x_j(t) = \sum_{i=j}^{j+t-1} ~ \xi_i, ~~~~j=1,2,...M-t+1 ~,
   \label{difftraj}
   \end{equation}
   where $x_j(t)$ represents the position of the trajectory 
   $j$ at time $t$.
   Scaling can be studied by direct evaluation of 
   the time behavior of the variance of the diffusion process: 
   \begin{equation}
   \sigma^2_x(t) \propto t^{2H}.
   \label{eq2}
   \end{equation}

   We note that this choice of trajectories is based on a window of size 
   $t$ the left side of which moves from the position $j=1$ to the 
   position $M-t+1$. The DFA rests on a much smaller number of 
   non-overlapping windows, whose left side is located at the positions 
   $1, t+1, 2t+1 ....$, and so on. For any of these non-overlapping 
   windows the DFA considers only the difference between the actual 
   sequence value and a local trend \cite{stanley3}. 
   The DE method uses, on the contrary, 
   the overlapping windows of Eq.(\ref{difftraj}). 
   This method of analysis, shown in action here for the first
   time on DNA sequences, is derived from that recently applied to 
   the analysis of time series of sociological interest \cite{nicola}, 
   and more details on it are given in ref.\cite{preparation}. 
Here we limit ourselves to explaining the motivation for the choice 
  of the overlapping windows of Eq. (\ref{difftraj}). In addition to increasing the 
  statistical accuracy of the analysis, the use of overlapping windows 
  is the same prescription as that dictated, at least in principle, by 
  the rules for the calculation of the Kolmogorov-Sinai entropy 
  \cite{beck,dorfman}. The DE shares with the KS the use of the Shannon entropy 
  indicator, as we shall see later, and also the same prescription to 
  convert one single trajectory in a large set of distinct 
  trajectories. The DE uses these trajectories to determine the scaling 
  of the diffusion process that is generated by the spreading of these 
  trajectories. The KS evaluates instead the rate of the entropy 
  increase associated to this spreading \cite{baranger}. If this spreading is 
  independent of biases, the DE determines the scaling associated to 
  this spreading without requiring de-trending, since the scaling is 
  determined by the entropy increase and this is virtually independent 
  of biases.   

   To evaluate the Shannon entropy of the diffusion process 
   at time $t$ we partition the x-axis into cells of size 
   $\epsilon =1$, and we define $S(t)$ as:
   \begin{equation}
   S(t) = - \sum_{i} p_{i}(t) ln [p_{i}(t)],
   \label{entropy} 
   \end{equation}
   where $p_{i}(t)$ is the probability that $x$ 
   can be found in the $i$-th cell at time $t$: 
   \begin{equation} 
   p_{i}(t) \equiv \frac{N_{i}(t)}{(M-t+1)} ,
   \label{probability}
   \end{equation} 
   and $N_{i}(t)$ is the number of trajectories found in the 
   cell $i$ at a given time $t$.
   The connection between
   $S(t)$ and scaling becomes evident in the continuous approximation, where the
   trajectories of the DNA walk of 
   eq.(\ref{difftraj}) are described by the continuous equation of motion: 
   \begin{equation}
   \frac{dx}{dt} = \xi(t) .
   \label{equationofmotion}
   \end{equation}
   Here $\xi(t)$ is the dichotomous variable assuming the values $+1$ 
   and $-1$, and $t$ is thought of as a continuous time. 
   In this case the Shannon entropy reads
   $ S(t) = - \int_{-\infty}^{\infty} dx \, p(x,t) ln [p(x,t)].$
   We assume:
   \begin{equation}
   p(x,t) = \frac{1}{t^{\delta(t)}} \, F\left( \frac{x}{t^{\delta(t)}}\right).
   \label{generalizedscaling} 
   \end{equation}
   This is a generalization of the ordinary scaling assumptions that can 
   be recovered by setting $\delta(t)$ equal to the time independent 
   scaling parameter $H$. 
   For the sake of simplicity we keep the 
   ordinary assumption of a fixed form of statistics, expressed by the 
   analytical form of the coefficient $A$ defined in 
    Eqs.(\ref{keyrelation}) and (\ref{ainthecontinuouscase}). 
   Using Eq.(\ref{generalizedscaling}),
   after a simple algebra, we get for the entropy: 
   \begin{equation}
   S(t) = A + \delta(t) ~ ln (t) ,
   \label{keyrelation}
   \end{equation}
   where
   \begin{equation}
   A \equiv -\int_{-\infty}^{\infty} dy \, F(y) \, ln [F(y)] .
   \label{ainthecontinuouscase}
   \end{equation}
   The diffusion entropy is a linear function 
   of the logarithm of $t$, with a slope equal to $\delta(t)$, 
   and this makes the slope measurement equivalent to 
   the scaling detection. 

   Let us now consider the two following possibilities:

   \noindent

   1) If $\xi(t)$ is an uncorrelated dichotomous variable, 
   $F(y)$ has a Gaussian form: 
   \begin{equation}
   F_{Gauss}(y) = \frac{exp\left(-\frac{y^{2}}{2 \sigma^{2}}\right)}{\sqrt{2\pi
   \sigma^{2}}},
   \label{gaussianform}
   \end{equation} 
   and then the diffusion entropy of Eq.(\ref{keyrelation}) reads 
   \begin{equation}
   S(t) = \frac{1}{2} \left[1 + \ln(2 \pi \sigma^{2})\right] + 
   \frac{1}{2} \ln(t) .
   \label{linearincrease1}
   \end{equation} 

   \noindent

   2) If, instead, $\xi(t)$ has the power-law correlation function 
  $\Phi_{\xi}(t) \sim 1/t^{\mu}$, with $0< \beta < 1$, the 
  distribution density of sojourn times in one of the two states $+1$                          
  or $-1$, $\Psi_{\xi}(t)$, is known \cite{allegro} to get the form
  $\Psi_{\xi}(t) \sim 1/t^{\beta}$, with $\mu = \beta + 2$. This implies a divergent second 
  moment and consequently \cite{mario} the $F(y)$ getting the form of a stable 
  L\'{e}vy distribution, thereby yielding:

   \begin{equation}
   S(t) = A_{Levy} + \frac{1}{\mu -1} \ln(t) .
   \label{linearincrease2}
   \end{equation} 
   For both cases we expect $S(t)$ to be a linear function of $ln(t)$, with slope
   $\delta = 0.5$ and $\delta = 1/(\mu-1)$, in
   the uncorrelated and correlated case, respectively. We note that uncorrelated
   Gaussian cases exist \cite{paolophdthesis},
   where $\delta =(4-\mu)/2$.


   We are now ready to consider the applications to the two DNA sequences. 
   In Fig. 1a we show that the DE analysis of the non-coding sequence 
   HUMTCRADCV results in a scaling changing with time, 
   and correlated diffusion shows up at both the short-time 
   and the long-time scale. 
   This is pointed out by means of two straight lines 
   of different slopes: the scaling in the short-time 
   regime $\delta=0.615$ coincides exactly with the value 
   found by means the DFA analysis \cite{stanley3}, 
   while the real asymptotic scaling is $\delta=0.565$ 
   corresponding to $\mu=2.77$ (see eq.(\ref{linearincrease2})).

   \begin{figure}[h]
   \epsfig{file=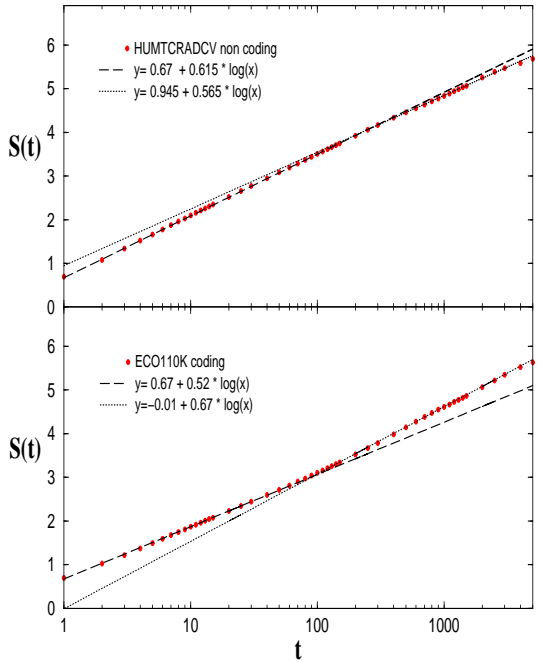, height=7cm,width=9cm,angle=-90}
   \caption{The diffusion entropy analysis for the two DNA sequences 
   results in a scaling changing with time. 
   For the HUMTCRADCV, the non-coding cromosomal fragment, 
   the slope of the straight line is $\delta=0.615$ at short-time regime, 
   and $\delta=0.565$ at long-time regime. 
   For ECO110K, the coding genomic fragment, 
   slopes are $\delta=0.52$ at short-time regime 
   and $\delta=0.67$ at long-time regime 
   }
   \end{figure}

   In Fig. 1b we consider the more delicate problem of a coding
   sequence: for ECO110K we observe at short time a slope $\delta=0.52$, very
   close to that of ordinary random walk, and at long-time a correlated
   diffusion with $\delta=0.67$, corresponding to $\mu=2.5$. 
   We note that the authors of Ref. \cite{stanley3} using the DFA 
   find in the short-time regime an uncorrelated diffusion with 
   $\delta_{V}=0.51$ in agreement with the DE, 
   and in the long-time regime a scaling $\delta_{V} = 0.75$, 
   which apparently conflicts with the finding of the DE method, 
   yielding $\delta = 0.67$. 
   Note that the symbol $\delta_{V}$, with $V$ standing for variance, 
   refers to the scaling detected by means of the DFA, which is in fact 
   based on the variance measurement. Actually, we can prove that this 
   apparent conflict yields a strong support to the main finding of our 
   paper, that the DE method reveals the long-range correlations and 
   the true asymptotic scaling of both coding and non-coding sequences.

   In order to do so, we model a DNA sequence by adopting the Copying 
   Mistaken Map (CMM) of Ref. \cite{barbi}. 
   As pointed out more recently 
   \cite{buiattino}, this model is equivalent to the 
   Generalized L\'{e}vy Walk (GLW) \cite{stanley2}. 
   The GLW, in turn, fits very well the observation made by the 
   authors of Ref. \cite{stanley3} that the transition to super-diffusion 
   in the long-time region is a manifestation of random walk patches 
   with bias. 
   The CMM corresponds to a picture where Nature builds up the 
   real DNA sequence, either coding or non-coding, by 
   using two different sequences. 
   The former is a Random Sequence (RS) equivalent to assigning to 
   any site the value +1 or -1 with equal probability. 
   The latter sequence, on the contrary, is highly correlated and is 
   obtained as follows. 
   First of all, a sequence of integer numbers $l > 0$ 
   is drawn, with the inverse power law distribution: 
   \begin{equation}\label{powerlaw}
   p(l)=\frac{C}{(T+l)^{\mu}}, ~~ 2 < \mu < 3  ~.
   \end{equation}
   Any drawing corresponds to fixing the length of a 
   sequence of patches. To any patch is then assigned a sign, either +1 
   or -1, by tossing a coin. 
   This prescription is virtually the same as that adopted 
   to build up the symbolic sequence of Ref. \cite{luigi}, 
   and corresponds to the intermittent condition of 
   the Manneville map \cite{gaspard}. 
   We call this correlated sequence Intermittent Randomness Sequence (IRS). 
   As shown in refs. \cite{allegro,mario}, the diffusion process 
   generated by the IRS is a L\'{e}vy diffusion. 
   %
   %
   According to the CMM, Nature builds up the real DNA sequence by 
   adopting for any site of the real sequence the nucleotide 
   occupying the same site in the RS, with probability $p_{R}$, 
   or the corresponding one of the IRS with probability 
   $p_{L} = 1 - p_{R}$. 
   The same prescription is used for modeling both the 
   coding and non-coding DNA sequences, 
   the only difference being in $p_{R}$, i.e. in the 
   percentage of correlated to uncorrelated component: 
   in particular the condition $p_{R} \gg p_{L}$ is valid for the 
   coding DNA. 
   The L\'{e}vy diffusion is faster than ordinary diffusion, 
   and therefore is expected to become predominant, 
   and so ostensible at long times, even when $p_{R} \gg p_{L}$. 
   Of course, upon increase of $p_{R}$ L\'{e}vy 
   statistics become ostensible at longer and longer times.
   As shown in Fig.2, the DE of HUMTCRADCV and ECO110K is perfectly 
   reproduced by a CMM with $\mu=2.77$ and $\mu=2.5$, respectively. 
   For the coding sequence $p_{R}=0.943$, i.e. the random component 
   is predominant, while for the non-coding sequence $p_{R}=0.560$. 
   It is worth to notice that with such values of $p_{R}$ the CMM also 
   accounts for the correct slope of $S(t)$ vs. $ln(t)$ 
   in the short-time regime. 

   Finally, we want to illustrate an important property of the DE method. 
   The DE detects the real scaling of 
   the distribution $\delta$, 
   rather than the second moment scaling $\delta_{V}$. 
   The two scaling values are identical only in the Gaussian case. 
   In the L\'{e}vy case they are related \cite{allegro} the one to the other by:                   
   \begin{equation}
   \label{eq14}
   \delta = \frac{1}{3 - 2 \delta_{V}} 
   \end{equation}
   We see that in the case of the non-coding sequence the DE yields an 
   asymptotic scaling which is slightly smaller than the short-time 
   scaling. This corresponds to the transition from the short-time 
   Gaussian condition to the long-time L\'{e}vy condition, namely, to the 
   transition from $\delta = \delta_{V} = 0.61$ at short time to the value 
   $\delta = 1/(\mu - 1) = 0.565$ of the L\'{e}vy regime, with delta related now 
   to $\delta_{V} = 0.61$ by Eq. (\ref{eq14}). In the coding case we see that the 
   scaling detected by the DE method is $\delta = 0.67$ that again is 
   related to $\delta_{V} = 0.75$ through Eq. (\ref{eq14}).

   \begin{figure}[h]
   \epsfig{file=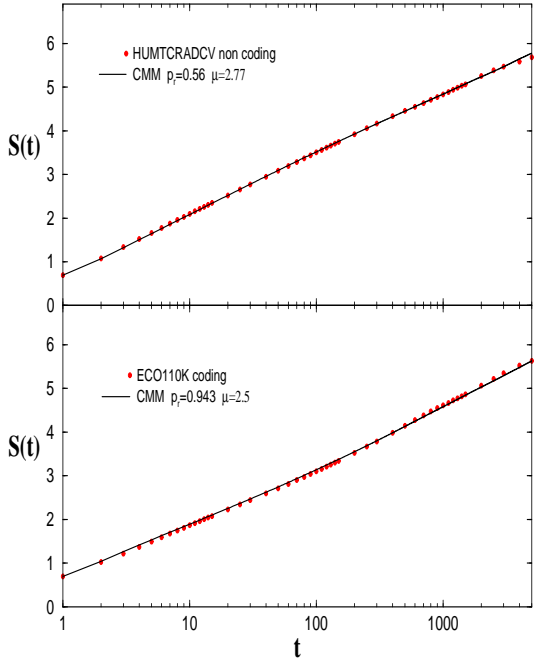, height=7cm,width=9cm,angle=-90}
   \caption{CMM simulation of the two DNA sequences. 
   Fig. 2a shows the comparison between the DE analysis of 
   HUMTCRADCV and an artificial sequence corresponding to the 
   CMM model with $p_R=0.56$, $T=0.43$, $\mu=2.77$.
   Fig. 2b shows the comparison between the DE analysis of 
   ECO110K and an artificial sequence corresponding 
   to the CMM model with $p_R=0.943$, $T=45$, $\mu=2.5$.
   }
   \end{figure}

   In conclusion, this paper affords two important results. 
   It proves that the DE method is a very reliable 
   technique that detects the real scaling, and the real 
   scaling does not coincide in general with that given by the DFA. 
   The second result is that the joint use of the DE and DFA 
   makes it possible to prove that the CMM, or the GLW, 
   which is totally equivalent to the 
   CMM \cite{buiattino}, 
   accounts for both coding and non-coding sequences. 
   All this strengthens the idea that both non-coding and coding DNA 
   sequences yield in the long-time limit an evident 
   manifestation of long-range correlations, 
   and confirms the claims of Ref. \cite{buiattino}, 
   where the non-Gaussian nature of the long-time regime was 
   interpreted as a sign of the L\'{e}vy 
   character of this region. 
   The L\'{e}vy nature of the long-time statistics is 
   now made compelling by the use of the DE, a method of statistical 
   analysis so accurate as to perceive the difference between 
   L\'{e}vy and Gauss scaling.


   \end{document}